\documentclass[11pt,a4paper]{article}
\usepackage[T1]{fontenc}
\usepackage[a4paper,top=2cm,bottom=2cm,left=3cm,right=3cm,marginparwidth=1.75cm]{geometry}
\usepackage{amsmath}
\usepackage{graphicx}
\usepackage{hyperref}
\usepackage{enumerate}
\usepackage[affil-it]{authblk}

\makeatletter
\renewcommand{\section}{\@startsection {section}{1}{\z@}%
              {24pt}{12pt} {\large\scshape\bfseries}}
\renewcommand{\subsection}{\@startsection {subsection}{2}{\z@}%
             {12pt}{12pt}  {\itshape\bfseries}}
\usepackage{apacite}
\usepackage{natbib}

\title{\bfseries \normalsize A Fundamental Analysis of the Impact on Traffic Assignment \\by Toll System of Electric Road System}
\author[1]{Wataru Nakanishi}
\author[2]{Noriko Kaneko}
\affil[1]{Associate professor, Department of Geosciences and Civil Engineering, Kanazawa University, Japan}
\affil[2]{Independent scholar (MEng), Japan}
\date{\vspace{-5ex}}
\begin{document}
\maketitle
\section*{Short summary}\small
Electric road system (ERS) is expected to make electric vehicles (EVs) more popular as EVs with Dynamic Wireless Power Transfer (DWPT) system can be charged while driving on ERS.
Although some studies dealt with ERS implementation, its toll system has not been explored yet. 
This paper aims at a fundamental analysis on impact of ERS toll system on a traffic assignment. 
We conduct assignments on a simple network where two vehicle types (EVs with DWPT and others) are co-existing. 
The results under two toll systems showed some undesirable situations, such as total travel time was not minimised, total charged volume was not optimised, and ERS was not utilised. 
The occurrence of them depended on the ratio of EVs, battery level, value of electricity, and toll price. 
The difficulty to control such situations by toll price was discussed as the battery level and value of electricity may vary over time.

\textbf{Keywords}: Dynamic wireless power transfer, Electrification and decarbonization of transport, Pricing and capacity optimization, Route choice, Social optimum.
\section{Introduction}
Electric vehicles (EVs) are being introduced to reduce CO$_2$ emissions in transport division.
Electric road system (ERS) is a future road system that EVs can be charged while driving on the road \citep{ERS2017SWE}.
ERS is expected to make EVs more popular and useful because it can save time by not standing still to be able to charge and increase their cruising distances.
EVs can utilise ERS if they have Dynamic Wireless Power Transfer (DWPT) system.
Hereafter, we call EVs with DWPT system as DWPT-EVs, and other vehicles including non-EVs as OTHER-Vs.

Related studies have dealt with 
optimal location of ERS implementation on a network
from both electricity perspective \citep{10081503} and transportation perspective \citep{riemann2015optimal, Liu17}.
Also, \cite{9936617} considered the interaction between electricity and transportation network to minimize the total cost related to ERS installation and operation. 
However, researches on traffic assignment problem under ERS environment is limited to \cite{SHI2022119619}, which solved a lane assignment.
Moreover, the impact of toll system on ERS to traffic assignment has not been analyzed yet. 
Toll system is mentioned merely as a list \citep{Bernecker}, a business model for a financial analysis \citep{HADDAD2022113275}, and indicators to be measured \citep{gustavsson2021research}.
Toll-free and road tolls are considered, and the latter has various types such as depending on vehicle type, time of utilization, and geographical locations.
In this paper, we deal with toll-free and fixed-toll systems as a preliminary analysis.

It is natural to assume a near-future situation where some of the vehicles on the roads are DWPT-EVs and the rest are OTHER-Vs, as in \cite{SHI2022119619}.
Also, some of the roads (or lanes) are ERS-implemented, and the rests are not.
A traffic assignment under this situation is similar to that with bus lanes and toll roads.
However, considering ERS and DWPT-EVs is more complicated in various aspects.
For example, due to the positive utility of charging, some DWPT-EVs may use ERS even though it is congested, which does not contribute to solving the environmental issues.
In addition, if optimal tolling to minimise total travel time (TTT) is applied, charged volume is not necessarily maximised.
Furthermore, electricity prices in the coming society are expected to be dynamic pricing with large fluctuations \citep{VoE2006US}.
This must affect the DWPT-EVs' choice whether to use ERS or not.

Based on the above, the aim of this paper is to examine the differences in the realized traffic assignment patterns depending on the toll system and the value of variables and parameters such as toll price, ratio of DWPT-EVs to all vehicles (DWPT-ratio), value of time (VoT), and value of charged electricity (VoE).
Also, some related values such as TTT, total amount of charged volume (TCV), and revenue for these patterns are calculated.
A simple network and simple utility functions for traffic assignment are assumed so that the realized traffic flow can be derived without conducting simulations. 
We illustrate that the assignment includes some complicated and problematic issues even in simple cases.
Also, we demonstrate through a numerical example that such patterns are not exceptional.
\section{Methodology}
\subsection{Assumption}
\begin{figure}[tb]
 \centering
 \includegraphics[scale=0.5]{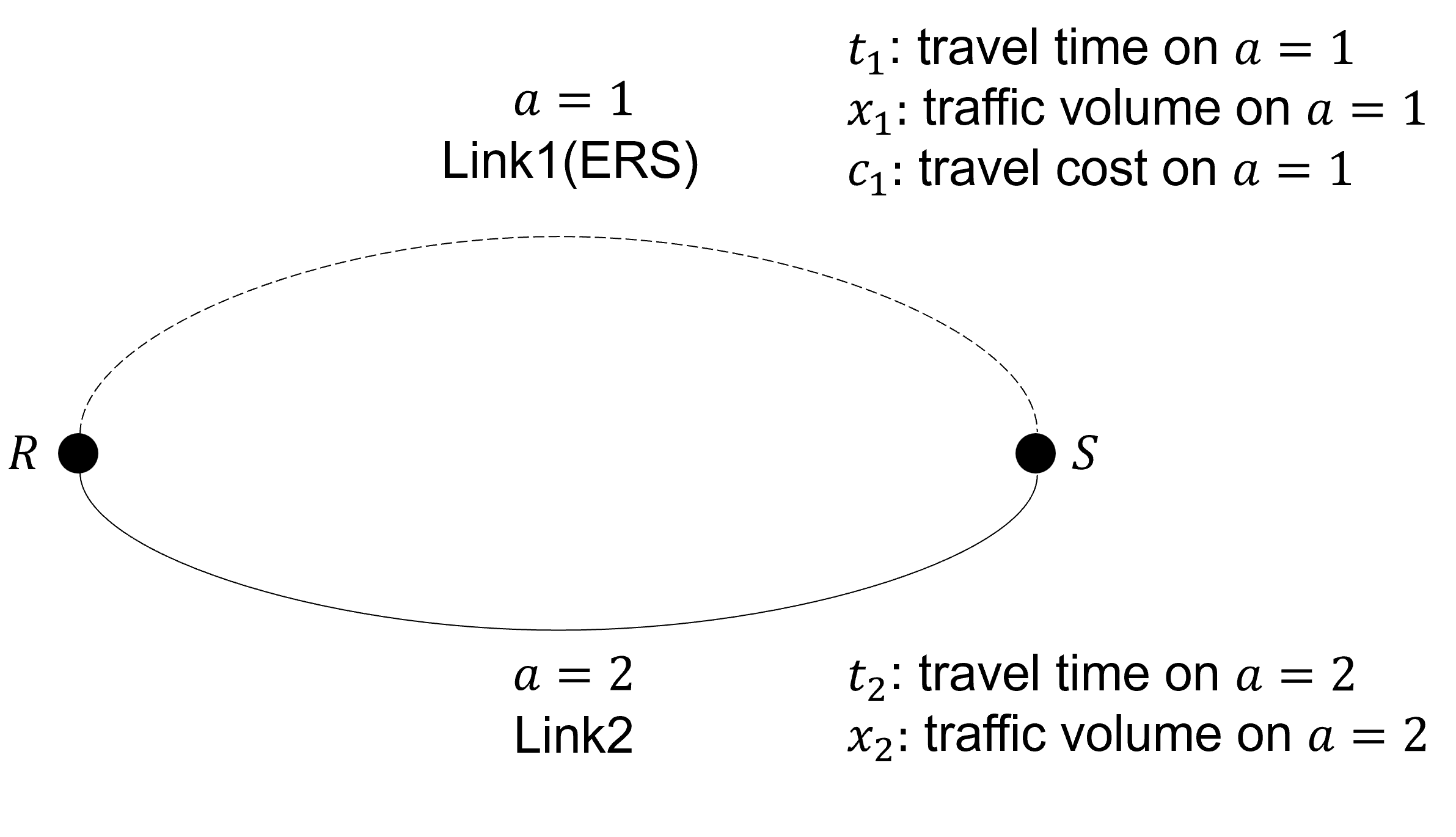} \\
 \vskip\baselineskip
 \caption{Network setting}
 \label{fig:network}
\end{figure}

A network consists of two nodes $(R, S)$ and two links $a=1, 2$.
There is one OD-pair from node $R$ to node $S$ (Fig.(\ref{fig:network})).
Link 1 ($a=1$) has ERS over the entire length while link 2 ($a=2$) doesn't have it.
Other properties of these links are the same.

$\mathcal D$ and $\mathcal O$ represent a set of DWPT-EVs and OTHER-Vs, respectively.
Total number of vehicles is denoted as $N$ and DWPT-ratio as $r,\ 0<r<1$;
the number of DWPT-EVs is $rN$.
DWPT-EV's traffic volume on $a$ is denoted as $x_a^D$, and Other-V's traffic volume on $a$ as $x_a^O$.

Then, some properties of DWPT-EVs are assumed as follows.
\begin{itemize}
    \item Battery capacity 
     is represented in units of electric energy (e.g., [kWh]), which is the product of electric power and time.
    \item The amount of battery consumed by a DWPT while driving is calculated by dividing the distance traveled (e.g., [km]) by the electric cost consumption (e.g., [km/kWh]).
    The amount a DWPT can charge while driving on the ERS is calculated by multiplying the travel time, $t$ (e.g., [h]), by the output of the ERS, $W$ (e.g., [kW]).
    \item State of charge (SoC) is defined as the ratio of the remaining battery level to the battery capacity.
    \item For simplicity, $i \in \mathcal D$ is always charged while driving on ERS.
\end{itemize}

\subsection{Model}
Each vehicle deterministically chooses one of the links (i.e., $a=1$ or $2$) to maximize its utility
as in a deterministic user equilibrium assignment. 
The utility functions of a vehicle 
are described by the weighted sum of the following three components.

\begin{enumerate}[i)]
    \item Travel time disutility. \\
    Travel time on link $a$, $t_a$, is described by a monotonically increasing function of traffic volume on link $a$, $x_a$ (e.g., The Bureau of Public Roads (BPR) function).
       
    \item Travel cost disutility.\\
    Travel cost on link $a$ is denoted as $c_a$.
    Only $c_1$ in fixed-toll system is defined in this paper.

    \item Charging utility.\\
    This is only applied to DWPT-EVs.
    The more electicity a DWPT-EV $i$ can charge, the larger this utility is.
    We denote SoC of $i$ at $R$ as $s_i$ and assume $0<s_i<1$ for simplicity.
    Then, we model this utility as $(1/s_i-1)$.
\end{enumerate}

\subsection{Formulation A -- toll-free system}
In this system, all vehicles use ERS for free.
This means that ERS is built and operated by taxes.
The utility functions are defined as Eqs.(\ref{eq:V_free-1}--\ref{eq:V_free-4}).

\begin{align}
V_{1,i}^D &= {\beta}_{t} t_{1} +\beta_{s} ( \frac{1}{s_i}-1 ) \label{eq:V_free-1}\\
V_{2,i}^D &= {\beta}_{t} t_{2}\label{eq:V_free-2}\\
V_{1,j}^O &= {\beta}_{t} t_{1}\label{eq:V_free-3}\\
V_{2,j}^O &= {\beta}_{t} t_{2}\label{eq:V_free-4}
\end{align}

Here, $V_{a,i}^D$ and $V_{a,j}^O$ represent the utility of $i \in \mathcal D$ and $j \in \mathcal O $ to choose link $a$, respectively.
$\beta_t$ and $\beta_s$ are the parameters of travel time disutility and charging utility, respectively.
$\beta_t<0$ and $\beta_s>0$ are assumed. 

\subsection{Formulation B -- fixed-toll system}
In this system, vehicles can use ERS for a fixed toll of $c_{1}=C$ (constant).
In particular, we consider that only $i \in \mathcal D$ uses $a=1$ should pay the toll.
This is the simplest implementation of the policy that any vehicles should pay about their ``fee for electricity'' \citep{gustavsson2021research}.
The utility functions are defined as Eqs.(\ref{eq:V_toll_2-1-1}--\ref{eq:V_toll_2-1-4}).
\begin{align}
        V_{1,i}^D &= {\beta}_{t}t_{1} + \beta_{c}C + \beta_{s}(\frac{1}{s_i}-1) \label{eq:V_toll_2-1-1} \\
        V_{2,i}^D &= {\beta}_{t}t_{2}\label{eq:V_toll_2-1-2} \\    
        V_{1,j}^O &= {\beta}_{t} t_{1} \label{eq:V_toll_2-1-3}\\
        V_{2,j}^O &= {\beta}_{t} t_{2}\label{eq:V_toll_2-1-4} 
\end{align}

Here, $\beta_c$ is the parameter of travel cost disutility.
$\beta_c<0$ is assumed. 
Also, $\beta_t/\beta_c$ and $-\beta_s/\beta_c$ represents VoT and VoE, respectively. 
\section{Results and discussion}
Traffic volume of each link and each vehicle type, TTT, TCV, and revenue are calculated under the systems and utility functions.
Also, the effects of DWPT-ratio $r$ and toll price $C$ are discussed.
To that end, all possible assignment patterns and the conditions for them are investigated for each system.

The results are broadly classified depending on whether DWPT-ratio $r<0.5$ or $r\geq0.5$.
In concrete, $x_1=x_2$ (hence, $t_1=t_2$) is always realized when $r<0.5$, namely, when the majority of vehicles are OTHER-Vs.
This is because OTHER-Vs can always achieve $x_1=x_2$ regardless of the DWPT-EVs' choice.
Moreover, if $x_1>x_2$, then $t_1>t_2$ and $V_{1,j}^O<V_{2,j}^O$ (and vice versa), which means $j \in \mathcal O$ has no motivation to realize other than $x_1=x_2$.

Note that $r<0.5$ is more realistic based on the current situation.
Nonetheless, if an assignment of multiple-origin-multiple-destination pair on a complex network is considered, links on some path alternatives might be mainly occupied by the DWPT-EVs.
Thus, investigating the case of $r\geq0.5$, which includes more complicated assignment patterns, will be also meaningful for subsequent research.

\subsection{Result A -- Toll-free system}
\begin{enumerate}[{A-}i.:]
    \item when $r<0.5$. \\
    As mentioned above, $x_{1}=x_{2}$ and $t_1=t_2$ 
    hold.
    All $i \in \mathcal D$ choose $a=1$ ($x_1^D=rN$) since $V_{1,i}^D-V_{2,i}^D = \beta_{s} ({1}/{s_i}-1)>0$.
    Also, $0.5N$ of OTHER-Vs choose $a=2$ and the rest choose $a=1$.
    
    In this case, 
    \begin{itemize}
        \item TTT is calculated as $x_1 t_1 + x_2 t_2 = Nt_1$ and is minimised.
        In conventional transportation network analyses, minimizing TTT is the most popular measure to determine social optimum (SO).
        In this sense, we refer to an assignment as ``conventional SO'' if TTT is minimised.
        \item TCV is calculated as $rNWt_1$. 
        Once traffic volume (and hence, travel time) is given and fixed, this is the possible maximum charged volume.
        We refer to such assignments as ``ERS-Optimum''.
        As $i$ cannot be charged beyond the battery capacity in the acutal situation, the precise calculation should be carried out in future work.
        \item Revenue is 0.
    \end{itemize}
    
    \item when $r \geq 0.5$. \\
    In this case, $x_1^D \geq 0.5N$.
    This can be checked by supposing the case $x_1^D<0.5N$, which means $x_1=x_2$ and both DWPT-EVs and OTHER-Vs choose both $a=1$ and $2$. 
    Although details are omitted due to space limitations, this is not a stable state.
    In a stable state, all $j \in \mathcal O$ should choose $a=2$ and $i \in \mathcal D$ chooses the link so that the travel time disutility and the charging utility are balanced.
    From Eq.(\ref{eq:V_free-1}), $i \in \mathcal D$ chooses $a=1$ in increasing order of $s_i$.
    Then, a threshold of $s_i$ that DWPT-EVs with smaller SoC than this value choose $a=1$ exists.
    We denote this SoC as $s_{thres}$ and the number of DWPT-EVs with $s_i < s_{thres}$ as $N_{thres}$.

    In this case, 
    \begin{itemize}
        \item TTT is calculated as $x_1 t_1 + x_2 t_2 = N_{thres}t_1 + (N-N_{thres}) t_2 = N_{thres}(t_1-t_2)+Nt_2$.
        This is minimised at the special case of $t_1 = t_2$, which means $N_{thres}=0.5N$ (i.e., $x_1=x_2$).
        Otherwise, generally, this case is not conventional SO.
        \item TCV is calculated as $N_{thres}Wt_1$, and this is ERS-optimum.
        Nonetheless, TCV is maximised at the special case of $N_{thres}=rN$ (i.e., all $i \in \mathcal D$ choose $a=1$) and not otherwise.
        \item Revenue is 0.
    \end{itemize}
\end{enumerate}

\subsection{Result B -- Fixed-toll system}
\begin{enumerate}[{B-}i.:]
    \item when $r<0.5$.\\
    The assignment of $j \in \mathcal{O}$ is uniquely determined to achieve $x_1=x_2$ once the assignment of $i \in \mathcal D$ is determined.
    Thus, we first consider the assignment of DWPT-EVs.
    The following three patterns are possible.
    \begin{enumerate}[{B-i.}(a)]
        \item Only $a=1$ is chosen by all $i \in \mathcal D$.\\
        This pattern occurs when even a DWPT-EV with maximum $s_i$ will pay the toll and charge its battery (i.e., $s_{\max}<s_{thres}$ where $s_{\max}$ represents the maximum $s_i$).
        Naturally, $x_2^O=0.5N$ and $x_1^O=(0.5-r)N$.
        Note that $t_1=t_2$, from Eqs.(\ref{eq:V_toll_2-1-1}-\ref{eq:V_toll_2-1-2}), the cost condition for this pattern is
        \begin{align}
        C < -\frac{\beta_s}{\beta_c}(\frac{1}{s_{\max}}-1) .
        \label{b1a}
        \end{align}
        
        \item Only $a=2$ is chosen by all $i \in \mathcal D$.\\
        This pattern occurs when even a DWPT-EV with minimum $s_i$ will not pay the toll and not charge its battery (i.e., $s_{thres}\leq s_{\min}$ where $s_{\min}$ represents the minimum $s_i$).
        Naturally, $x_1^O=0.5N$ and $x_2^O=(0.5-r)N$.
        Similarly, the cost condition for this pattern is
        \begin{align}
        C \geq -\frac{\beta_s}{\beta_c}(\frac{1}{s_{\min}}-1) .
        \end{align}
        
        \item Both $a=1$ and $2$ are chosen by $i \in \mathcal D$.\\
        This pattern occurs when some DWPT-EVs will pay the toll and charge its battery while others will not (i.e., $s_{\min}<s_{thres} \leq s_{\max}$).
        $j \in \mathcal O$ chooses both $a=1$ and $2$ to realize $x_1=x_2$.
        Therefore, in this pattern, both DWPT-EVs and OTHER-Vs choose both links.
        Similarly, the cost condition for this pattern is
        \begin{align}
        -\frac{\beta_s}{\beta_c}(\frac{1}{s_{\min}}-1) \leq C < -\frac{\beta_s}{\beta_c}(\frac{1}{s_{\max}}-1) .
        \label{b1c}
        \end{align}
    \end{enumerate}
    In this case,
    \begin{itemize}
        \item TTT is $Nt_1$, which means Conventional SO.
        \item TCV is calculated as $N_{thres}Wt_1$.
        This is maximised at the special case of $N_{thres}=rN$: pattern (a).
        Also, ERS-optimum is only achieved in pattern (a).
        Otherwise, TCV is not maximised and the assignments are not ERS-Optimum.
        In particular, TCV becomes 0 in (b).
        \item Revenue is calculated as $N_{thres}C$.
    \end{itemize}

    \item when $r \geq 0.5$.\\
    As for $i \in \mathcal D$, the same three assignment patterns as B-i. can be done as follows.
    Corresponding assignment patterns for $j \in \mathcal O$ are also described. 
    \begin{enumerate}[{B-ii.}(a)]
        \item Only $a=1$ is chosen by all $i \in \mathcal D$.\\
        Here, all $j \in \mathcal O$ choose $a=2$.
        At this point, $x_1=rN > x_2=(1-rN)$ and hence, $t_1>t_2$.
        Then, if no DWPT-EV has a motivation to change its link choice under this $t_1$ and $t_2$ (i.e., a DWPT-EV with $s_{\max}$ still chooses $a=1$), this is a stable assignment.
        From Eqs.(\ref{eq:V_toll_2-1-1}-\ref{eq:V_toll_2-1-2}), the cost condition for this pattern is
        \begin{align}
        C < \frac{\beta_t}{\beta_c}(t_1-t_2)-\frac{\beta_s}{\beta_c}(\frac{1}{s_{\max}}-1) .
        \label{b2a}
        \end{align}
        
        \item Only $a=2$ is chosen by all $i \in \mathcal D$.\\
        Here, all $j \in \mathcal O$ choose $a=1$.
        At this point, $x_1=(1-rN) < x_2=rN$ and hence, $t_1<t_2$.
        Then, if no DWPT-EV has a motivation to change its link choice under this $t_1$ and $t_2$ (i.e., a DWPT-EV with $s_{\min}$ still chooses $a=2$), this is a stable assignment.
        Similarly, the cost condition for this pattern is
        \begin{align}
        C \geq \frac{\beta_t}{\beta_c}(t_1-t_2)-\frac{\beta_s}{\beta_c}(\frac{1}{s_{\min}}-1) . 
        \end{align}
        
        \item Both $a=1$ and $2$ are chosen by $i \in \mathcal D$.\\
        (c1) Suppose $x_1=x_2$ after all $j \in \mathcal O$ choose the links. 
        Then, this is similar to B-i.(c).
        Both DWPT-EVs and OTHER-Vs choose both links except the special cases of exactly $0.5N$ of DWPT-EVs choose $a=1$ or $2$.\\
        (c2) Suppose $x_1 > x_2$ after all $j \in \mathcal O$ choose the links.
        This occurs only if all $j \in \mathcal O$ choose $a=2$ and $N_{thres}>0.5N$ holds.\\
        (c3) Suppose $x_1 < x_2$ after all $j \in \mathcal O$ choose the links.
        This occurs only if all $j \in \mathcal O$ choose $a=1$ and $(rN-N_{thres})>0.5N$ holds.\\
        The cost condition common to (c1-3) is
        \begin{align} 
            \frac{\beta_t}{\beta_c}(t_1-t_2)-\frac{\beta_s}{\beta_c}(\frac{1}{s_{\min}}-1)  \leq C <  \frac{\beta_t}{\beta_c}(t_1-t_2)-\frac{\beta_s}{\beta_c}(\frac{1}{s_{\max}}-1) .
            \label{b2c}
        \end{align}
    \end{enumerate}
    In this case,
    \begin{itemize}
        \item TTT is only minimised in (c1) and not otherwise.
        \item TCV is calculated as $N_{thres}Wt_1$.
        This is maximised at $N_{thres}=rN$ in (a) and not otherwise.
        In particular, TCV becomes 0 in (b).
        ERS-optimum is achieved in pattern (a), a special case of (c1), and (c2).
        \item Revenue is $N_{thres}C$.
    \end{itemize}
\end{enumerate}

\subsection{Discussion}
When $r<0.5$, toll-free system (A-i.) is less problematic except for the beneficiary-pay perspective.
This assignment is social optimum in the conventional meaning, and is maximising charged volume, if the policy that ERS construction and operating cost are covered by taxes is acceptable.
On the other hand, the biggest problem in fixed-toll system (B-i.) is that ERS may not be utilised by DWPT-EVs depending on the price $C$.
Some DWPT-EVs avoid ERS to avoid paying toll (B-i.(c)).
Furthermore, in the worst situation, no DWPT-EV uses ERS and no revenue is obtained (B-i.(b)).
Even in the simple example of this paper, a beneficiary-pay-oriented system may lead to such meaningless and undesirable situations. 
Numerical examples are shown in the next subsection.
Of course, if we could specify an appropriate price $C$, the resulting assignment has no problem (Eq.(\ref{b1a})).
However, this seems to be completely difficult because the upper bound of $C$ is a function of $s_{\max}$, $\beta_s$ and $\beta_c$.
First of all, we cannot expect $s_{\max}$ to be constant or stable, either for within-day or day-to-day variations.
In addition, $\beta_s$ is not always stable and is assumed to vary over time in accordance with electricity prices \citep{VoE2006US}.

When $r\geq0.5$, the biggest problem common to A-ii. and B-ii. is that TTT is not generally minimised.
If spread of ERS and DWPT-EVs results in increased TTT, they are counterproductive to CO$_2$ emission reductions.
Therefore, the case $r\geq0.5$ is more difficult than the case $r<0.5$ in that the toll system must be designed in such a way that TTT does not increase inadvertently.
Also, the same problem as $r<0.5$ that ERS is not fully utilised exists.
Of course, theoretically, $C$ could be determined to minimise TTT or to achieve ERS-optimum.
However, this is difficult because of the same reason as $r<0.5$.

\subsection{Numerical example of case B-i.}
Finally, a numerical example is shown by taking case B-i. as an example.
Some values assuming Japanese current market are determined (Table \ref{tab:hensuu}).
Travel time is calculated by BPR function, $t_a = t_{a0}(1+\alpha {x_a}/{Q_a})^\beta$. 

\begin{table*}[hbtp]
 \caption{Variables for numerical examples} 
 \label{tab:hensuu}
\begin{center}
\begin{tabular}{cll}
\hline
variable & description & value\\\hline
$N$ &total number of vehicles &1000\\
$r$ &ratio of DWPT-EV &0.2\\
$s_i$ & SoC of DWPT-EV &$0.1\leq s_i \leq 0.9 $\\
 & (uniformly distributed between 0.1 and 0.9)& \\
$t_{a0}$ &free flow travel time (same for $a=1,2$) &10 [min]\\
 & (e.g., 60 [km/h], 10 [km] length road.)& \\
$Q_a$ & capacity of link (same for $a=1,2$) &500\\
$W$&output of the ERS &30 [kW]\\
$\alpha$ & parameter of BPR function &0.15\\
$\beta$ & parameter of BPR function &4\\  \hline
\end{tabular}
\end{center}
\end{table*}

\begin{figure}[hbtp]
 \centering
 \includegraphics[scale=0.5]{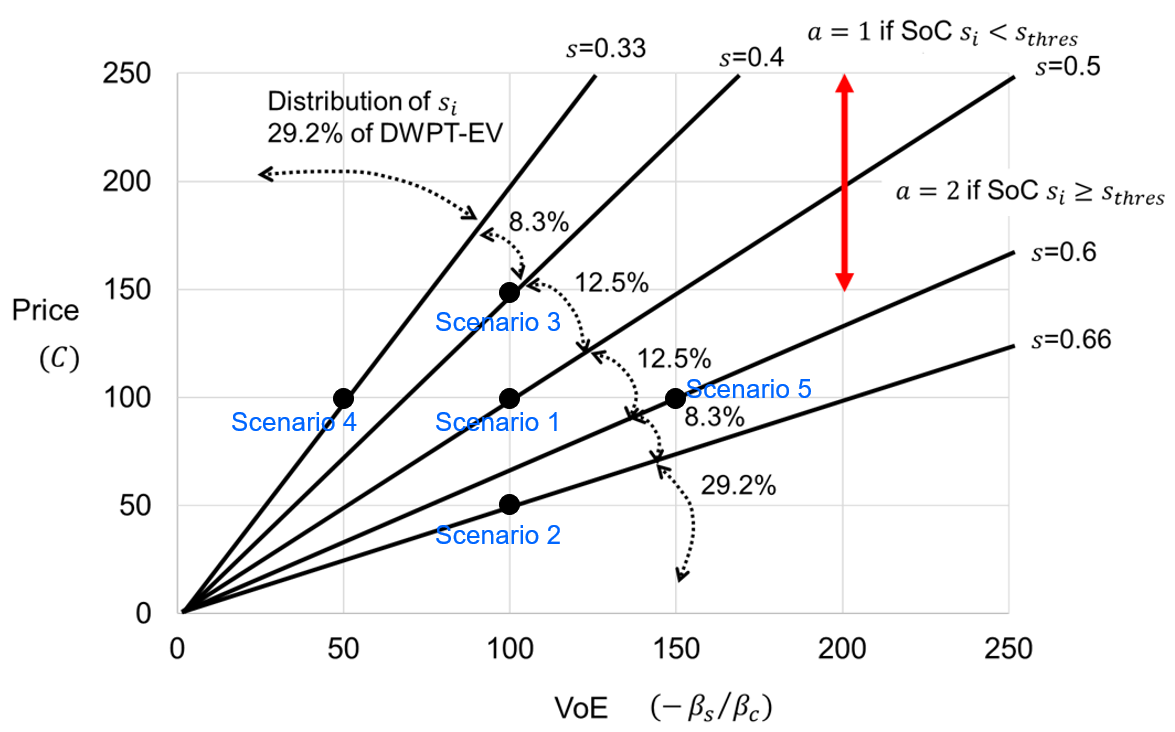} \\
 \vskip\baselineskip
 \caption{Relationship among VoE, $C$, and $s_{thres}$}
 \label{fig:voe}
\end{figure}

The results of five scenarios are shown in Table \ref{tab:results}.
They differ in VoE and $C$:
Scenario 1 is the base case; Scenarios 2 to 5 are those with VoE or $C$ multiplied by 1.5 or halved.
VoE $=100$ and and the price $C=100$ [JPY] are assumed by standard VoT (50 [JPY/min]) and the output of quick charger (around 120 [kW]).
Since the charging utility is defined in a non-linear way, VoE cannot be directly interpreted as the price of electricity, but it can vary in such degrees.
Fig. \ref{fig:voe} is the illustration of Eq.(\ref{b1c}).
$s_{thres}$ is the value of $s_i$ when $V_{1,i}^D = V_{2,i}^D$ (Eq.(\ref{eq:V_toll_2-1-1}-\ref{eq:V_toll_2-1-2})) holds, and is calculated from VoE and $C$.

As a result, changes in both price $C$ (from Scenarios 1--3) and VoE (Scenarios 1, 4, 5) made a difference of several tens of per cent of DWPT-EVs in ERS utilisation. 
Also, the significant changes in ERS utilisation for different SoC distributions can also be easily confirmed by considering the scenarios where the values on Fig. \ref{fig:voe} are changed.
For example, if $s_{\min}>0.4$ in Scenario 3, no DWPT-EV choose $a=1$, which is the worst pattern.

\begin{table*}[hbtp]
 \caption{Results of traffic assignment} 
 \label{tab:results}
\begin{center}
\begin{tabular}{c r r r r r }
\hline
Scenario &  1 & 2& 3 &  4&  5\\\hline
VoE &        100 & 100 & 100 & 50 & 150    \\ 
$C$ &        100 & 50 & 150 & 100 & 100    \\ 
$s_{thres}$&      0.50 & 0.66 & 0.40 & 0.33 & 0.60    \\ \hline
$x_1^D$   &      100.0 & 141.7 & 75.0 & 58.3 & 125.0   \\ 
$x_2^D$ &        100.0 & 58.3 & 125.0 & 141.7 & 75.0    \\ 
$x_1^O$ &        400.0 & 358.3 & 425.0 & 441.7 & 375.0    \\ 
$x_2^O$ &        400.0 & 441.7 & 375.0 & 358.3 & 425.0    \\ 
$x_1(=x_2)$ &        500.0 & 500.0 & 500.0 & 500.0 & 500.0    \\ 
$t_1(=t_2)$ &       11.5 & 11.5 & 11.5 & 11.5 & 11.5    \\ \hline
TTT  &        2300 & 2300 & 2300 & 2300 & 2300    \\ 
TCV &        575.3 & 814.6 & 431.3 & 335.5 & 718.8    \\ 
Revenue &        10000.7 & 7083.7 & 11251.0 & 5834.0 & 12500.7   \\ \hline
\end{tabular}
\end{center}
\end{table*}

\section{Conclusions}
In this study, the effect of toll system of ERS on traffic assignment was analysed.
It was shown that even simple situations entails problems that cannot be ignored.
In concrete, the assignment results differ greatly depending on the toll system, price and VoE.
Also, TTT, TCV and the degree of ERS utilisation differ significantly.
Therefore, in addition to the location decision from both electricity and transportation viewpoints, the aspect of toll system should be considered to decide the location to introduce ERS.
The problem becomes more difficult when time variation of VoE is large and majority of vehicles is DWPT-EV. 

Future directions are as follows.
Firstly, it is necessary to sort out the desirable state of traffic assignment in terms of TTT, TCV and ERS construction and operational cost burdens, as they differ from each other.
Secondly, considering the real world problem, a simulation study on larger networks and multiple OD pairs is important. 
From a modelling viewpoint, other toll systems, such as those based on the amount of charge or time of use, should also be investigated. 
Furthermore, in DWPT-EV decision-making, improving the charging utility function and adding an error term to utility functions are possible.

\section*{Acknowledgements}
This research was partially supported by JSPS KAKENHI 22H01610 and 23K17551.

\bibliography{references}

\end{document}